\documentclass[fleqn,twoside]{article}
\usepackage{gc,epsf}
\usepackage [russian]{babel}
\heads{Boris N. Frolov}
      {On Foundations of Poincar\'{e}--gauged Theory of Gravity}

\begin{document}
\twocolumn[
\Arthead{10}{2004}{1-2 (37-38)}{116}{120}

\Title{ON FOUNDATIONS OF POINCAR\'{E}-GAUGE THEORY OF GRAVITY}

\Author{Boris N. Frolov\foom 1}
       {Moscow State Pedagogical University, Department of Physics for Nature Sciences,\\
        Krasnoprudnaya 14, Moscow 107140, Russian Federation}

\Abstract
    {Foundations of the Poincar\'{e}-gauge theory of gravity are developed.
It is shown that the Poincar\'{e}-gauge field consists of two components:
the translational gauge field ($t$-field), which is generated by the
energy-momentum current of external fields, and the rotational gauge
field ($r$-field), which is generated by the sum of the angular and spin
momentum currents of external fields. Therefore, a physical field generated
by the angular momentum of a rotating mass should exist.}

]
\email 1 {frolovbn@mail.ru}

\section{Introduction}

In the second part of XX century, after the formula\-tion of the idea
of the gauge treatment of gravity in the pioneering works of R. Utiyama
\cite{Ut} and D. Ivanenko with collaborators \cite{BrIvSok},
the Poincar\'{e}-gauge theory of gravity (PGTG) was developed,
see \cite{Kib}--\cite{Frbook} and the references therein, and also
\cite{Sar} on the modern mathematical approach to gauge theory.
But the gauge approach to the construction of a modern
gravitational theory contains many unclear and unsolved problems
up to now. In this context, a problem of a great interest
consists in the possible existence of a still unknown physical field
that can be generated by rotating masses as a source. In
particular, such a possibility follows from the theorem on the
sources of gauge fields \cite{Fr1}, \cite{Frdis},
\cite{Frbook}, which states that the gauge field is generated by
the Noether invariant corresponding to the Lie group, which
introduces this gauge field by the localization procedure. In this
sense, the gauge field should be generated not only by the
energy-momentum tensor, but also by a sum of angular and spin
momentum currents as a source. Therefore a physical field
generated by the angular momentum of a rotating mass should exist
\cite{Frdis}--\cite{Frbook}.
\par
The corresponding field equations of PGTG were derived in
\cite{Frdis}, \cite{Frbook} as a consequence of the general gauge
field theory for groups connected with space-time
transformations. We have shown that, under the localization
procedure, the Lorenz subgroup of the Poincar\'{e} group introduces
a rotational gauge field ($r$-field), which is generated by the
sum of the angular and spin momentum currents of external fields.
The subgroup of translations introduces a translational gauge
field  ($t$-field), which is generated by the energy-momentum
current of the external fields. These field equations are equivalent
to the equations of PGTG in a usual form, derived by variation
of the Lagrangian with respect to the tetrads and connections,
the tetrads, curvature and torsion being constructed with
the help of both the $r$- and $t$-fields.

\section{The Noether theorem and the local invariance principle}

We start with flat Minkowski space $M_4$  with Cartesian coordinates
$x^{a}$ ($a =1,2,3,4$) and the metric $\breve g$ with the basis
$\vec{e}_{a} = \partial_{a} = (\partial /\partial x^{a})$.
The fundamental group of $M_4$ is Poinca\-r\'{e} group ${\cal P}_4$
(inhomogeneous Lorentz group),
\begin{eqnarray}
\delta x^{a} = \omega^{m}I_{m}\!^{a}\!_{b} x^{b} + a^{a} =\omega^{z}X^{a}_{z}=
- \omega^{z} \hat{M}_{z} x^{a}\; ,\nonumber \\
\{\omega^{z}\} = \{\omega^{m},\,a^k\}\;,\qquad
\hat{M}_z = \{\hat{M}_m,\,\hat{M}_k\}\;,\nonumber \\
\hat{M}_{m} = - I_{m}\!^{a}\!_{b} x^{b}\frac{\partial}{\partial x^{a}} \;, \quad
\hat{M}_k = P_{k} = - \frac{\partial}{\partial x^{k}}\; . \nonumber
\end{eqnarray}
Here $\{\omega^{z}\} = \{\omega^{m},\,a^k\}$ are the infinitesimal para\-me\-ters
of ${\cal P}_4$, and we have introduced the abbre\-vi\-a\-tions for the rotations
and translations,
\[
X^a_z = \{X^a_m,\, X^a_k \}\;, \quad X^a_m = I_{m}\!^{a}\!_{b} x^{b}\;,
\quad X^a_k = \delta^a_k\; .
\]
\par
Let us introduce the curvilinear system of coordinates
$\{x^{\mu}\}= \{x^{\mu}(x^{a})\}$  on $M_4$:
\begin{eqnarray}
&& dx^{a} =\hat{h}^{a}\!_{\mu} dx^{\mu}\,, \;\; \hat {h}^{a}\!_{\mu}
= \vec{e}_\mu (x^a) = \frac{\partial }{\partial x^{\mu}}(x^a)\,, \nonumber\\
&& ds^{2} = \hat{g}_{\mu\nu} dx^{\mu} dx^{\nu}\;, \quad
g_{\mu\nu} = g_{ab} \hat{h}^{a}\!_{\mu} \hat{h}^{b}\!_{\nu}\, ,\nonumber\\
&& \hat{g} = \mbox{det}(\hat{g}_{\mu\nu}) = \mbox{det}(g_{ab})\hat{h}^2 \, ,\nonumber\\
&& \hat{h} = \mbox{det} (\hat{h}^{a}\!_{\mu} ) = \sqrt{\mid \hat{g}\mid}\,.  \nonumber
\end{eqnarray}
\par
The action integral is invariant under the Poincar\'{e} group ${\cal P}_4$,
\begin{eqnarray}
&& J = \int\,(dx)\,\sqrt{\mid \hat{g}\mid}\, L(Q^A\,,\,P_{k}Q^A)\;,\nonumber\\
&& P_k = - \hat{h}^{\mu}\!_{k}\partial_\mu\;, \qquad
{\cal L} =\sqrt{\mid \hat{g}\mid}\, L = \hat{h} L\; .\nonumber
\end{eqnarray}
Noether first theorem in a curvilinear system of coordinates yields the following,
\begin{eqnarray}
&& 0 = \int \, (d x) \Biggl (\frac{\delta {\cal L}}{\delta Q^A}
\bar{\delta} Q^A  \nonumber\\
&& + \partial_\mu \left ({\cal L} \hat{h}^{\mu}
\!_{k}\delta x^k - \hat{h}^{\mu}\!_{k}\frac{\partial{\cal L}}
{\partial P_k Q^A}\bar{\delta} Q^A\right )\Biggl )\; . \nonumber
\end{eqnarray}
Here $\bar\delta$  denotes variation of the form of the field. For example,
for the field $Q^A$ we have,
\[
\bar{\delta} Q^A = \delta Q^A -\delta x^\mu \partial_\mu Q^A \;.
\]
The field equation of the field $Q^A$  is fulfilled,
\[
0 = \frac{\delta {\cal L}}{\delta Q^A} = \frac{\partial {\cal L}}{\partial Q^A}
+ \partial_\mu \left (\frac{\partial{\cal L}}{\partial P_k Q^A}
\hat{h}^{\mu}\!_{k}\right ) \; .
\]
\par
The result of Noether theorem can be represen\-ted as follows,
\[
0 = \int\,(d x)\, \hat{h}\hat{h}^{\mu}\!_{k}\, \partial_\mu
(a^l  t^{k}\!_{l} + \omega^m M^{k}\!_{m} )\; ,
\]
where the following expressions for the energy-momentum $t^{k}\!_{l}$
and the full momentum $M^{k}\!_{m}$  (angular momentum plus spin momentum
$J^{k}\!_{m}$) tensors are introduced:
\begin{eqnarray}
t^{k}\!_{l} = L\delta^{k}_{l} - \frac{\partial L}{\partial P_k Q^A}P_lQ^A\; ,\nonumber\\
M^{k}\!_{m} = J^{k}\!_{m} + I_{m}\!^{l}\!_{b}x^{b}t^{k}\!_{l}\; , \nonumber\\
J^{k}\!_{m} = - \frac{\partial L}{\partial P_k Q^A} I_{m}\!^{A}\!_{B} Q^B \; .\nonumber
\end{eqnarray}
Noether's theorem yields the conservation laws for the energy-momentum and full momentum,
\[
P_k\,t^{k}\!_{l} =0\;, \qquad P_k\,M^{k}\!_{m} = 0\; .
\]
\par
Now we shall localize the Poincar\'{e} group ${\cal P}_4$, whose parameters $\omega^z$
become arbitrary functions of the coordinates on $M_4$. The theory is based on four Postulates
\cite{Fr1}, \cite{Frdis}, \cite{Frbook}.
\par
Postulate 1 ({\em The local invariance principle}) The action integral
\[
J = \int\,(d x)\,{\cal L}(Q^A\,,\,P_k Q^A\,;\,A^R_a\,,\,P_k A^R_a )\;,
\]
where the Lagrangian density ${\cal L}$ describes the field $Q^A$, its interaction
with the additional gauge field $A^R_a$ and also the free gauge field $A^R_a$,
is invariant under the action of the localized group ${\cal P}_4 (x)$, the gauge field
being transformed as follows,
\begin{equation}
\delta A^R_a = U^{R}_{za}\,\omega^z + S^{R\mu}_{za} \partial_\mu \omega^z \;, \label{eq:varA}
\end{equation}
where $U$ and $S$ are unknown matrices.
\par
Postulate 2 ({\em The stationary action principle}):
\begin{equation}
\frac{\delta {\cal L}}{\delta Q^A} = 0\; , \qquad
\frac{\delta {\cal L}}{\delta A^R_a} = 0\; . \label{eq:varurQA}
\end{equation}
\par
Postulate 3 ({\em The existence of a free gauge field}). The full Lagrangian
density ${\cal L}$ of the physical system has the following structure,
\begin{eqnarray}
{\cal L} = {\cal L}_0 + {\cal L}_Q\; , \quad {\cal L}_0 = {\cal L}_0
(A^R_a\,,\,P_k A^R_a)\; , \nonumber \\
\frac{\partial {\cal L}_0}{\partial Q^A}= 0\;,
\quad \frac{\partial {\cal L}_0}{\partial P_k Q^A} = 0\; , \nonumber
\end{eqnarray}
where ${\cal L}_0$ is the Lagrangian density of the free gauge field, and
${\cal L}_Q$ is the Lagrangian density of the interaction of gauge field
with external fields.
\par
Postulate 4 ({\em The minimal gauge interaction principle}):
\[
\frac{\partial {\cal L}_Q}{\partial P_k A^R_a} = 0\; .
\]
\par
Noether's second theorem for our Lagrangian density and the group
${\cal P}_4$ yields the following,
\begin{eqnarray}
0 = \int\, (d x) \left (\frac{\delta {\cal L}}
{\delta Q^A}\bar{\delta}Q^A + \frac{\delta {\cal L}}{\delta A^R_a}\bar{\delta}
A^R_a \right )   \nonumber \\
+ \int (d x)\, \partial_\mu \Biggl ({\cal L} \hat{h}^{\mu}\!_{k}\delta x^k
- \hat{h}^{\mu}\!_{k} \frac{\partial {\cal L}}{\partial P_k Q^A}
\bar{\delta} Q^A \nonumber\\
-\hat{h}^{\mu}\!_{k} \frac{\partial {\cal L}}
{\partial P_k A^R_a}\bar{\delta} A^R_a \Biggr )\, . \nonumber
\end{eqnarray}
Here, in $\delta x^k,\; \bar{\delta} Q^A,\; \bar{\delta} A^R_a$ we have arbitrary
functions $\delta\omega^z (x), \; \partial_\mu\delta\omega^z (x), \;
\partial_\mu\partial_\nu\delta\omega^z (x)$, and the coefficients
before them are equal to zero:
\begin{eqnarray}
 \partial_\mu (\hat{h}\hat{h}^{\mu}\!_{k}\,\Theta^{k}\!_{z} ) +
(U^{R}_{za} + X^l_z P_l A^R_a )\frac{\delta {\cal L}}{\delta A^R_a} = 0\; .
\label{eq:ft1} \\
 \hat{h}\hat{h}^{\mu}\!_{k}\,\Theta^{k}\!_{z}
+ \partial_\nu {\cal M}^{\nu\mu}\!_{z} +
\frac{\delta {\cal L}}{\delta A^R_a} S^{R\mu}_{za}= 0 \; ,\label{eq:ft2} \\
{\cal M}^{(\nu\mu)}\!_{z} = 0\; , \label{eq:ft3}
\end{eqnarray}
where the following notations are introduced,
\begin{eqnarray}
\lefteqn {
\hat{h} \Theta^{k}\!_{z} = {\cal L}X^{k}_{z} - (I_{m}\!^{A}\!_{B}Q^B +
X^{l}_{z}P_l Q^A )\frac{\partial {\cal L}}{\partial P_k Q^A} } \nonumber\\
&&- (U^{R}_{za} +  X^l_z P_l A^R_a )\frac{\partial {\cal L}}{\partial P_k A^R_a} \; ,
\nonumber \\ 
&& {\cal M}^{\nu\mu}\!_{z} = \hat{h}\!^{\nu}\!_{k}\,\frac{\partial {\cal L}}
{\partial P_k A^R_a} S^{R\mu}_{za}\; . \nonumber
\end{eqnarray}
If Eqs (\ref{eq:varurQA}) for the gauge field are valid, then Eqs
(\ref{eq:ft1})--(\ref{eq:ft3}) are simplified:
\begin{eqnarray}
&&\partial_\mu (\hat{h} \hat{h}^{\mu}\!_{k}\, \Theta^{k}\!_{z} ) = 0\; , \nonumber\\
&& \hat{h} \hat{h}\!^{\mu}\!_{k}\Theta^{k}\!_{z} +
\partial_\nu {\cal M}^{\nu\mu}\!_{z} = 0\; ,\;\;
{\cal M}^{(\nu\mu)}\!_{z} = 0\; . \nonumber
\end{eqnarray}

\section{Structure of the Lagrangian densities ${\cal L}_Q$ and ${\cal L}_0$}

We introduce the differential operator $M_R$:
\begin{eqnarray}
&& M_R = \{ M_m\,,\, \hat{M}_k \}\; ,\label{eq:MR}\\
&& M_m = I_m + \hat{M}_m\; , \quad \hat{M}_k = P_k \; , \nonumber
\end{eqnarray}
and represent the gauge field $A^R_a$ in two compo\-nents,
$A^R_a = \{A^{m}_{a}\,,\,A^{k}_{a} \}$, where $A^k_a$ describes the translational
component of the gauge field ($t$-field), and $A^m_a$ describes the rotational
component of the gauge field ($r$-field).
\par
{\bf Theorem 1} (B.N. Frolov, 1999, 2003). {\em There exists a gauge field $A^R_a$  with
transformational struc\-ture of Postulate 1 under the localized Poincar\'{e} group
${\cal P}_4 (x)$, and there are such matrix functions $Z$, $U$ and $S$ of the gauge field
that the Lagrangian density
\begin{equation}
{\cal L}_Q = h L_Q (Q^A\,, D_a Q^A)\; , \qquad h = Z \hat{h}\; ,  \label{eq:LQ}
\end{equation}
satisfies Postulate 1, ${\cal L}_Q$ being constructed from $L(Q^A\,,P_k Q^A)$
by exchanging the operator $P_k$ with the  gauge derivative operator
\begin{equation}
D_a = - A^{R}_{a} M_R \; , \label{eq:AM}
\end{equation}
where the operator $M_R$ is given by (\ref{eq:MR}). The fol\-low\-ing representation
for the gauge $t$-field is valid,}
\begin{equation}
A_a^k = D_a x^k \; . \label{eq:BDx}
\end{equation}
\par
A proof of this Theorem \cite{Frdis}, \cite{Frbook} consists
in proving three prepositions. \newline
{\bf Preposition 1}. With the help of (\ref{eq:MR}), the gauge derivative (\ref{eq:AM})
can be represented as follows:
\begin{eqnarray}
 D_a Q^A &=& h^\mu\!_{a}\, \partial_\mu Q^A - A^m_a I_{m}\!^{A}\!_{B} Q^B \nonumber\\
& =& h^{\mu}\!_{a}D_\mu Q^A\;,  \label{eq:hD} \\
D_\mu Q^A &=& \partial_\mu Q^A - A^{m}\!_{\mu}I_{m}\!^{A}\!_{B} Q^B\; , \nonumber
\end{eqnarray}
where new quantities are introduced,
\begin{eqnarray}
&& Y^k_a = A^R_a\,X^k_R = A^k_a + A^m_a\, X_{m}^{k}\; , \label{eq:Y}\\
&& Z^a_k = (Y^{-1})^a_k\;, \qquad h^{\mu}\!_{a} = \hat{h}^{\mu}\!_{k}\, Y_a^k\;,\label{eq:Y1}\\
&& h^a\!_{\mu} = Z^a_k\, \hat{h}^{k}\!_{\mu}\;,
\qquad A^{m}\!_{\mu} = h^a\!_\mu \, A^{m}\!_{a}\; . \label{eq:hamu}
\end{eqnarray}
It is easy to verify with the help of (\ref{eq:hD}) and (\ref{eq:Y})--(\ref{eq:hamu})
that Eq. (\ref{eq:BDx}) is valid. \newline
{\bf Preposition 2}. We represent the Noether identi\-ties (\ref{eq:ft2}), (\ref{eq:ft3}) as
a set of differential equations for the unknown function  ${\cal L}_Q$, taking into account the
minimal gauge interaction principle (Postulate 4). The solvability conditions of the first of these
equations are satisfied, if the unknown matrix functions $Z$ and $S$ have the form
\begin{eqnarray}
 S^{n\mu}_{ma} = \delta^n_m h^{\mu}\!_{a} \, , \;\;  S^{n\mu}_{ka} = 0\, ,\;\;
S^{l\mu}_{ma} = 0\; , \label{eq:S}\\  S^{l\mu}_{ka} = \delta^l_k h^{\mu}\!_{a}\,,\;\;
Z = \mbox{det}(Z^a_k)\,, \label{eq:S1}\\  h = Z \hat{h} = \mbox{det}(Z^a_k)\,\mbox{det}
(\hat{h}^a\!_{\mu}) = \mbox{det}(h^a\!_{\mu}) \, . \label{eq:h}
\end{eqnarray}
{\bf Preposition 3}. Substituting the results of Prepositions 1 and 2 into
the set of equations (\ref{eq:ft1}), we shall see that this set of equations is
satisfied identically by the Lagrangian density (\ref{eq:LQ}) provided that the unknown
matrix function $U$ has the form
\begin{eqnarray}
&& U_{ma}^{n} = c_m\!^n\!_q A_{a}^{q} - I\!_m\!^b\!_a A_{b}^{n}\;,\label{eq:Un}\\
&& U_{ma}^{k} = I\!_m\!^k\!_l A_{a}^{l} - I\!_m\!^l\!_a A_{l}^{k} \; , \label{eq:Una}\\
&& U_{ka}^{n} = 0\;,\quad U_{ka}^{l} = - A_{a}^{n}I\!_n\!^l\!_k \; . \label{eq:Unz}
\end{eqnarray}
\par
{\bf Corollary}. Substituting (\ref{eq:S})--(\ref{eq:Unz})
into (\ref{eq:varA}), we get the transformational laws of the gauge fields
$A^R_a = \{A^{m}_{a}\,,\,A^{k}_{a} \}$ under the localized group ${\cal P}_4 (x)$:
\begin{eqnarray}
\delta A^m_a &=& \omega^{n}(x)\, c_{n}\!^m\!_q \,A_{a}^{q} -
\omega^{n}(x)\,I_n\!^b\!_a\, A_{b}^{m}  \nonumber\\
&& + h^{\mu}\!_{a}\,\partial_{\mu}\omega^{m}(x)  \nonumber \\
& =& D_a \omega^m - \omega^{n}(x)\, I_n\!^b\!_a\, A_{b}^{m}\; , \nonumber\\
\delta A_{a}^{k} &=& \omega^{m}(x)\, (I_m\!^k\!_l\, A_{a}^{l} - I_m\!^l\!_a\, A_{l}^{k})\nonumber\\
&&- A^m_a \,I_m\!^k\!_l \,a^l (x) + h^{\mu}\!_{a}\,\partial_{\mu} a^{k}(x) \nonumber \\
& = & D_{a} a^{k} +  \omega^{m}\,(I_m\!^k\!_l\, A_{a}^{l} - I_m\!^l\!_a\, A_{l}^{k})\;,\nonumber\\
\delta h^\mu\!_a &=& - \omega^{m}(x)\, I_m\!^b\!_a \,h^\mu\!_b + h^\nu\!_a\, \partial_\nu \delta x^\mu\; ,
\nonumber\\
\delta h^{a}\!_{\mu} &=& \omega^{m}(x)\, I_m\!^a\!_b \,h^{b}\!_{\mu} -
h^a\!_\nu \,\partial_\mu \delta x^\nu \; ,  \nonumber\\
\delta A^m\!_\mu &=& D_\mu \omega^m - A^m\!_\nu \,\partial_\mu \delta x^\nu \; .  \nonumber
\end{eqnarray}
\par
One of the main result of this corollary is that the tetrads $h^\mu\!_a$  and $h^{a}\!_{\mu}$
{\it are not the true gauge potentials} in contrast to the usually accepted opinion \cite{Kib},
\cite{Tr}--\cite{Bas}, \cite{He2}. In this connection it should be mentioned that D. Ivanenko
and G. Sardana\-shvi\-ly \cite{Iv-Sar} argued against ``the hypothetical iden\-ti\-fi\-ca\-tion
of tetrad gravitational fields and translation gauge potentials'' in PGTG.
\par
The structure of the gauge field Lagrangian density ${\cal L}_{0}$ is established
by the following theorem.
\par
{\bf Theorem 2} (B.N. Frolov, 1999, 2003). {\em The Lagrangian density
\begin{equation}
{\cal L}_{0} = hL_{0}(F^{m}\!_{ab},\,T^{c}\!_{ab})\; , \label{eq:L0FT}
\end{equation}
where $L_{0}$ is an arbitrary scalar function of the gauge field strengthes
$F^{m}\!_{ab}$ and $T^{c}\!_{ab}$,
\begin{eqnarray}
&&F^{m}\!_{ab} = 2h^\lambda\!_{ [a} \partial_{\mid\lambda\mid} A^m_{b]} +
C^c\!_{ab} A^m_c - c_n\!^m\!_q A^n_a A^q_b\,, \nonumber\\
&&T^{c}\!_{ab} = C^c\!_{ab} + 2I_n\!^c\!_{ [a} A^n_{b]}\,,\nonumber\\
&&C^c\!_{ab} = - 2h^c\!_\tau h^\lambda\!_{ [a}
\partial_{\mid\lambda\mid} h^\tau\!_{b]} = 2 h^\lambda\!_a h^\tau\!_b
\partial_{ [\lambda} h^c\!_{\tau ]}\,, \nonumber
\end{eqnarray}
satisfies the local invariance principle (Postulate 1).}

\section{Field equations of the gauge fields}

The gauge field equations are
\begin{eqnarray}
&& \frac{\delta{\cal L}_0}{\delta A_{a}^{k}} = - \frac{\partial {\cal L}_{Q}}
{\partial A_{a}^{k}} = h t^a\!_k\;, \label{eq:istAk}\\
&& h t^a\!_k = Z^a_l \left ({\cal L}_{Q}\delta^{l}_{k} -
\frac{\partial{\cal L}_{Q}}{\partial P_l Q^{A}} P_k Q^A \right )\;,\nonumber\\
&& \frac{\delta{\cal L}_0}{\delta A_{a}^{m}} = -\frac{\partial {\cal L}_{Q}}
{\partial A_{a}^{m}} = h(\hat{M}\!\,^a\!_{m} + J^a\!_{m} )\;, \label{eq:istAm}\\
&& h\hat{M}\!\,^a\!_{m} = I_m\!^k\!_b x^b\, (ht^a\!_k)\;, \nonumber\\
&& hJ^a\!_{m} = \frac{\partial{\cal L}_{Q}}{\partial D_a Q^{A}} I_m\!^A\!_B Q^B\;.\nonumber
\end{eqnarray}
\par
The consequence of (\ref{eq:istAk}) and (\ref{eq:istAm}) is the theorem.
\par
{\bf Theorem 3} (B.N. Frolov, 1963, 2003; Theorem on the source of a gauge field).
{\em The source of the gauge field, introduced by the localized group ${\cal P}_4 (x)$,
is the Noether current corresponding to the non-localized group ${\cal P}_4$.}
\par
In the geometrical interpretation of the theory, the quantities $h^a\!_\mu$ and $A^m\!_\mu$
become tetrad fields and a Lorenz connection, while the quantities
\begin{eqnarray}
F^m\!_{\mu\nu} = F^m\!_{ab}\,h^a\!_\mu h^b\!_\nu = 2\partial_{ [\mu}
A^m\!_{\nu ]} - c_n\!^m\!_q A^n\!_\mu A^q\!_\nu \; , \nonumber\\
T^c\!_{\mu\nu} = T^c\!_{ab}\, h^a\!_\mu h^b\!_\nu = 2\partial_{ [\mu}h^c\!_{\nu ]}
+ 2I_n\!^c\!_a h^a\!_{ [\mu} A^n\!_{\nu]}\; .\nonumber
\end{eqnarray}
become the curvature and torsion tensors, respecti\-ve\-ly. The following theorem is valid.
\par
{\bf Theorem 4} (B.N. Frolov, 1999, 2003). The set of gauge field equations
(\ref{eq:istAk}), (\ref{eq:istAm}) derived by variation with respect to the gauge
fields $\{A^{k}_{a}\,,\,A^{m}_{a}\}$ is equivalent to the set of field equations
derived by variation with respect to the fields $\{h^a\!_\mu\,,\,A^m\!_\mu \}$,
\begin{equation}
\frac{\delta{\cal L}_0}{\delta h^{a}\!_{\mu}} = -\frac{\delta{\cal L}_Q}{\delta h^{a}\!_{\mu}}\; ,
\quad \frac{\delta{\cal L}_0}{\delta A^m\!_{\mu}} = -\frac{\delta{\cal L}_Q}{\delta A^m\!_{\mu}}\;.
\label{eq:vurAmu}
\end{equation}
\par
The first of the gauge field equation (\ref{eq:vurAmu}) can be represented in the form,
\begin{eqnarray}
&& \partial_{\nu} \frac{\partial{\cal L}_0}{\partial T^{a}\!_{\nu\mu}} =
\frac{1}{2}h (t^{\mu}_{(0)a} + t^{\mu}_{(Q)a} )\; ,\label{eq:urTab} \\
&& ht^{\mu}_{(Q)a} = h^\mu\!_a {\cal L}_Q - \frac{\partial{\cal L}_Q}
{\partial D_\mu Q^{A}}D_a Q^A\; , \nonumber\\
&& ht^\mu_{(0)a} = h^\mu\!_a {\cal L}_0 - 2F^m\!_{a\nu}\frac{\partial{\cal L}_0}
{\partial F^m\!_{\mu\nu}}  \nonumber \\
&& - 2 T^c\!_{a\nu}\frac{\partial{\cal L}_0}{\partial T^c\!_{\mu\nu}} +
2A^m\!_\nu I_m\!^c\!_a \frac{\partial{\cal L}_0}{\partial T^{c}\!_{\mu\nu}}\;,\nonumber
\end{eqnarray}
and the second one can be represented as
\begin{eqnarray}
&& \partial_\nu \frac{\partial {\cal L}_0}{\partial F^{m}\!_{\nu\mu}} =
-\frac{1}{2}h (J_{(0)m}^{\mu} + J_{(Q)m}^{\mu} )\; , \label{eq:urFmn}\\
&& hJ_{(Q)m}^{\mu} = \frac{\partial {\cal L}_Q}{\partial A^m\!_\mu} =
\frac{\partial{\cal L}_Q}{\partial D_\mu Q^A} I_m\!^A\!_B Q^B\;, \nonumber\\
&& hJ_{(0)m}^{\mu} = \frac{\partial{\cal L}_0}{\partial A^m\!_\mu} = \nonumber\\
&& = 2\frac{\partial{\cal L}_0}{\partial F^n\!_{\mu\nu}}c_m\!^n\!_q A^q\!_\nu +
2\frac{\partial{\cal L}_0}{\partial T^c\!_{\mu\nu}}I_m\!^c\!_a h^a\!_\nu \;. \nonumber
\end{eqnarray}
\par
The field equations (\ref{eq:urTab}) and (\ref{eq:urFmn}) yield the conser\-vational
laws for the canonical energy-momentum tensor $t^{\mu}_{(Q)a}$ of the external field
supplemented by the energy-momentum tensor $t^{\mu}_{(0)a}$ of the free
gauge field, and for the spin current $J_{(Q)m}^{\mu}$  of the external field
supplemented by the spin current $J_{(0)m}^{\mu}$ of the free gauge field,
\begin{eqnarray}
&& \partial_{\mu} (ht_{(0)a}^{\mu} + ht_{(Q)a}^{\mu}) = 0\;, \nonumber\\
&& \partial_{\mu} (hJ_{(0)m}^{\mu} + hJ_{(Q)m}^{\mu}) = 0\; . \nonumber
\end{eqnarray}
\par
The field equations (\ref{eq:urTab}) and (\ref{eq:urFmn}) can be represented
in a geometric form:
\begin{eqnarray}
&& D_\nu \left (\frac{\partial{\cal L}_{0}}{\partial T^{a}\!_{\nu\mu}}\right )
+ F^{m}\!_{a\nu}\frac{\partial{\cal L}_{0}}{\partial F^{m}\!_{\mu\nu}} \nonumber\\
&& + T^c\!_{a\nu}\frac{\partial{\cal L}_{0}}{\partial T^{c}\!_{\mu\nu}}
   -\frac{1}{2}h^{\mu}\!_a {\cal L}_{0} = \frac{1}{2}ht^{\mu}_{(Q)a}\,,\label{eq:DTab}
\end{eqnarray}
\[
D_{\nu} \left (\frac{\partial{\cal L}_{0}}{\partial F^{m}\!_{\nu\mu}}\right ) =
- \frac{1}{2}h J_{(Q)m}^{\mu} +
\frac{\partial{\cal L}_{0}}{\partial T^{b}\!_{\nu\mu}}I_m\!^b\!_a h^a\!_\nu \; . \nonumber
\]
\par
If we have ${\cal L}_{0} = hL_{0}(F^{m}\!_{ab})$ instead of (\ref{eq:L0FT}),
then the field equation (\ref{eq:DTab}) is simplified and generalizes the Hilbert--Einstein
equation to arbi\-trary nonlinear Lagrangians \cite{Fr1},
\[
F^{m}\!_{a\nu}\frac{\partial{\cal L}_{0}}{\partial F^{m}\!_{\mu\nu}} -
\frac{1}{2}h^{\mu}\!_a {\cal L}_{0} = \frac{1}{2}ht^{\mu}_{(Q)a}\; .
\]
\section{Conclusions}

The main result of the Theorem on the source of the gauge field
(Theorem 4) is that the sources of gauge field in PGTG are not
only the energy-momentum and the spin momentum tensors as in the
Einstein--Cartan theory \cite{Kib}, \cite{Tr}, but also the angular
momentum tensor. The gauge $t$- and $r$-fields are generated
together by the energy-momentum, angular momen\-tum and spin
momen\-tum tensors \cite{Frdis}--\cite{Frbook}. Therefore in PGTG
rotating masses (such as galaxies, stars and planets), as well as
polarized media, should generate the $r$-field. In
particular, a gyroscope on Earth should change its weight
subject to changing the direction of rotation because of the
interaction with the rotating Earth. There exists some experi\-mental
evidence of such effects. In \cite{Hay}--\cite{Vol}, changes of
the weight of rotating bodies or polarized media have been
observed, which can be explained by interaction of
these bodied and media with the rotating Earth. To this end, one
may also refer the results of N. Kozyrev's experiments with
gyroscopes \cite{Koz} and the mysteri\-ous relation between angular
moments and masses of all material bodies in our Metagalaxy
\cite{Mur}, \cite{Sis}. In Russia, in the Research Institute of
Space System (NIIKS of M V Khru\-ni\-chev GKNP Center), some hopeful
results have been obtained in experiments, which has demon\-s\-tra\-ted a
possibility of using the weight decrease of a rotating mass for
constructing an engine that could move a body without any contacts
with other bodies and without ejection of any reactive mass \cite{Men}.
In \cite{Tuck}, the influence of the rotating Sun on the planet motion
via generation of torsion has been investigated.
\par
In General Relativity (GR) torsion vanishes, and one gets
only one field equation with the metric energy-momentum tensor as
a source. But nevertheless the effects of GR depend on both
the $t$- and $r$-fields. In particular, the Lense--Thirring effect
and the Kerr solution are induced by the $r$-field. The well-known
problem of constructing an external source for the Kerr
metric \cite{Lop}, \cite{Bur} may have a solution in considering
the angular momentum of the external medium as the source.
\par
In GR, the coupling constants of the $t$- and $r$-fields are
equal to the Einstein gravitational constant. But in PGTG  this
choice is not deter\-mined by the theory and the coupling constants
of the $t$-field and $r$-field are not necessarily equal to each other.
In PGTG these constants can have different values, which should be
found only from the experimental data.

\renewcommand{\refname}{References}

\end{document}